\documentclass[12pt]{article}
\usepackage{amsfonts,amssymb,epsfig,amsmath,mathtools}
\usepackage{color}

\usepackage{mathrsfs}
\usepackage{graphicx}
\usepackage{subcaption}

\renewcommand{\baselinestretch}{1.2}

\newcommand{\be}{\begin{eqnarray}}
\newcommand{\ee}{\end{eqnarray}}

\newcommand{\bn}{\begin{enumerate}}
\newcommand{\en}{\end{enumerate}}

\begin{document}

\makeatletter \@addtoreset{equation}{section} \makeatother
\renewcommand{\theequation}{\thesection.\arabic{equation}}
\renewcommand{\thefootnote}{\alph{footnote}}

\begin{titlepage}

\begin{center}
\hfill {\tt KIAS-P22052}
\\
\hfill 
\\

\vspace{2cm}

{\Large\bf The shape of non-graviton operators for $SU(2)$}


\vspace{2cm}

\renewcommand{\thefootnote}{\alph{footnote}}

{\large Sunjin Choi$^1$, Seok Kim$^2$, Eunwoo Lee$^2$ and
Jaemo Park$^3$}

\vspace{0.7cm}

\textit{$^1$School of Physics, Korea Institute for Advanced Study,\\
85 Hoegi-ro, Dongdaemun-gu, Seoul 02455, Republic of Korea}\\

\vspace{0.2cm}

\textit{$^2$Department of Physics and Astronomy \& Center for
Theoretical Physics,\\
Seoul National University, 1 Gwanak-ro, Gwanak-gu, Seoul 08826, Republic of Korea}\\

\vspace{0.2cm}

\textit{$^3$Department of Physics, Pohang University of Science and Technology
(POSTECH),\\
77 Cheongam-ro, Nam-gu, Pohang 37673, Gyeongsangbuk-do, Republic of Korea}

\vspace{0.7cm}

E-mails: {\tt sunjinchoi@kias.re.kr, seokkimseok@gmail.com\\
eunwoo42@snu.ac.kr, jaemo@postech.ac.kr
}

\end{center}

\vspace{1cm}

\begin{abstract}

The BPS spectrum of AdS/CFT exhibits multi-gravitons at low energies,
while having black hole states at higher energies. This can be studied
concretely in AdS$_5$/CFT$_4$ in terms of classical cohomologies, even in the
quantum regimes at finite $1/N$. Recently, Chang and Lin found a
threshold for non-graviton states in the $SU(2)$ maximal super-Yang-Mills
theory. We explicitly construct and present this threshold cohomology.

\end{abstract}

\end{titlepage}

\renewcommand{\thefootnote}{\arabic{footnote}}

\setcounter{footnote}{0}

\renewcommand{\baselinestretch}{1}



\section{Introduction and the setup}

Regular BPS black holes in AdS$_5\times S^5$ preserve $2$ real Killing spinors
\cite{Gutowski:2004ez},
and should be dual to the $\frac{1}{16}$-BPS states of the 4d $U(N)$ maximal
super-Yang-Mills theory on $S^3\times \mathbb{R}$. By the operator-state map of a CFT,
these correspond to the $\frac{1}{16}$-BPS local operators on $\mathbb{R}^4$. These
operators are defined and constructed as follows (we follow the conventions of \cite{Biswas:2006tj,Grant:2008sk}).
The $U(N)$ adjoint fields of the super-Yang-Mills theory are given by
\begin{eqnarray}
  \textrm{6 real scalars}&:&\Phi_{ij}={\textstyle \frac{1}{2}}\epsilon_{ijkl}\Phi^{kl}\ \
  \textrm{where}\ \ \Phi^{kl}\equiv(\Phi_{kl})^\ast\nonumber\\
  \textrm{fermions}&:&\Psi_{i\alpha}\ ,\ \ \overline{\Psi}^{i}_{\dot\alpha}\nonumber\\
  \textrm{gauge fields}&:&A_{\mu}\sim A_{\alpha\dot\beta}
\end{eqnarray}
where $i,j,\cdots=1,\cdots,4$ are the $SU(4)$ $R$-symmetry indices, and
$\alpha=\pm$, $\dot{\alpha}=\dot{\pm}$ are the $SU(2)_l$ and $SU(2)_r$ indices in
$SO(4)\cong SU(2)_l\times SU(2)_r$ rotation symmetry, respectively. The system has $16$ Poincare
supersymmetries $Q^i_\alpha$ and $\overline{Q}_{i\dot\alpha}$. It also has
$16$ conformal supersymmetries which, in the radially quantized setup, are Hermitian
conjugate to $Q$'s: $S_{i}^{\alpha}=(Q^{i}_{\alpha})^\dag$,
$\overline{S}^{i \dot\alpha}=(\overline{Q}_{i \dot\alpha})^\dag$.
The $\frac{1}{16}$-BPS operators are annihilated by $Q\equiv Q^4_-$ and
$S\equiv S_4^-\equiv Q^\dag$. $Q$ and $S$ satisfy
\begin{equation}\label{algebra}
  2\{Q,S\}=E-(R_1+R_2+R_3+J_1+J_2)\geq 0\ ,
\end{equation}
where $R_{1,2,3}$ are the Cartans of $SO(6)\cong SU(4)$, and
$J_{1,2}$ are the Cartans of $SO(4)$. The BPS operators are made of the
following set of fields and derivatives,
which individually saturate the last inequality of
(\ref{algebra}) (in the free theory limit):
\begin{equation}\label{BPS-letters}
  \bar\phi^m\equiv\Phi^{4m}\ ,\ \ \psi_m\equiv \Psi_{m+}\ ,\ \
  \bar\lambda_{\dot\alpha}\equiv\overline{\Psi}^4_{\dot\alpha}\ ,\ \
  f\equiv F_{++}\ ,\ \ D_{+\dot\alpha}\ .
\end{equation}
Here, $F_{++}$ is a component of the field strength
$F_{\alpha\beta}\sim F_{\mu\nu}(\sigma^{\mu\nu})_{\alpha\beta}$, and
$D_{+\dot{\alpha}}$ are two components of the covariant derivatives
$D_{\alpha\dot{\beta}}\sim (\sigma^\mu)_{\alpha\dot{\beta}}D_\mu$.
The charges carried by these fields and derivatives are given as follows.
($R_I$ charges of other fields $\bar\phi^m$, $\psi_m$
are obtained by obvious permutations.)
\begin{table}[h!]
\begin{center}
\begin{tabular}{c||c|c|c}
	\hline
    fields & $(R_1,R_2,R_3)$ & $(J_1,J_2)$ & $E$\\
	\hline $\bar\phi^1$ & (1,0,0) & (0,0) & $1$ \\
    \hline $\psi_1$ & $(-\frac{1}{2},\frac{1}{2},\frac{1}{2})$ &
    $(\frac{1}{2},\frac{1}{2})$ & $\frac{3}{2}$   \\
    \hline $\bar{\lambda}_{\dot{\pm}}$ & $(\frac{1}{2},\frac{1}{2},\frac{1}{2})$
    & $(\pm\frac{1}{2},\mp\frac{1}{2})$ & $\frac{3}{2}$ \\
	\hline $f$ & (0,0,0) & $(1,1)$ & $2$ \\
    \hline $D_{+\dot{+}}$, $D_{+\dot{-}}$ & $(0,0,0)$ &
    (1,0), (0,1) & $1$ \\
    \hline
\end{tabular}\caption{Charges carried by the BPS fields and derivatives}
\end{center}
\end{table}\\

\vskip -1.5cm

All gauge invariant operators made of (\ref{BPS-letters}) saturate
the inequality of (\ref{algebra}) at strictly zero coupling, thus providing
the BPS operators of the free Yang-Mills theory. At nonzero coupling,
the composite operators typically acquire nonzero anomalous dimensions
and many BPS operators in the free theory become non-BPS. Questions were
raised in \cite{Kinney:2005ej,Berkooz:2006wc,minwalla} concerning which BPS
operators in the free theory
remain BPS in the `weakly interacting' regime, and whether they can account for
the strongly coupled BPS spectrum captured by the dual black hole entropies.
At 1-loop level, the first part of the question can be rephrased as a
classical cohomology problem in $Q$, which satisfies $Q^2=0$. $Q$-actions
on the BPS letters are given by
\begin{eqnarray}\label{Q-action}
  &&[Q,\bar\phi^m]=0\ ,\ \ 
  \{Q,\psi_m\}=-ig_{\rm YM}\epsilon_{mnp}[\bar\phi^n,\bar\phi^p]\ ,
  \nonumber\\
  &&\{Q,\bar\lambda_{\dot\alpha}\}=0\ ,\ \ 
  [Q,f]=-ig_{\rm YM}[\psi_m,\bar\phi^m]\ ,\ \
  [Q,D_{+\dot\alpha}]=-ig_{\rm YM}[\bar\lambda_{\dot\alpha},\ \ \}\ .
\end{eqnarray}
The second part of the question, whether the BPS operators at 1-loop level
remain all the way to the strong coupling, remains a conjecture
without known evidence against it. See \cite{Chang:2022mjp}
for a non-renormalization theorem on this issue.

With the last conjecture assumed, we expect the following structures of
the large $N$ spectrum of cohomologies, from the AdS dual. At energies much lower
than $N$, the spectrum is expected to be saturated by supergravitons. This has been
tested in \cite{Janik:2007pm,Chang:2013fba}. At energies of order $N$, we expect less
operators than gravitons because of the `giant graviton' effects \cite{McGreevy:2000cw}. This is implemented in QFT by the trace relations of finite matrices. These giant
graviton states, at least those explicitly constructed so far, are still of `graviton type' (whose meaning will be stated below). Finally, beyond certain energy level
(whose threshold value is not clearly known), we expect to have new cohomologies not of
the graviton type. We expect new cohomologies because some of them should
account for the microstates of AdS black holes at energies of order $N^2$.
In particular, the index \cite{Kinney:2005ej} of this
QFT exhibits a large number of operators to account for the
Bekenstein-Hawking entropy of the dual black holes. A selection of studies which
gave increasing confidence about this fact is (among many): Euclidean quantum gravity
analysis \cite{Cabo-Bizet:2018ehj}; the saddle point analysis \cite{Choi:2018hmj,Choi:2021rxi}; the Bethe ansatz analysis \cite{Benini:2018ywd};
explicit enumeration of the index at finite but reasonably large $N$ and charges
\cite{Murthy:2020rbd,Agarwal:2020zwm}.

The graviton type BPS cohomologies are characterized as follows \cite{Grant:2008sk}.
We first consider the single trace operators made of scalars $\bar\phi^m\equiv(X,Y,Z)$.
Since $[\bar\phi^m,\bar\phi^n]$ are $Q$-exact from (\ref{Q-action}), the scalars are
regarded as commuting operators inside the trace, e.g. ${\rm tr}(X^2YZ)\sim{\rm tr}(XYXZ)
\sim {\rm tr}(X^2ZY)$. Then, one acts the $9$ fermionic elements of $SU(1,2|3)$
commuting with $Q,S$ to obtain certain descendants \cite{Kinney:2005ej}.
Then one further acts ordinary derivatives $\partial_{+\dot{\alpha}}$ on these single
trace operators to obtain conformal descendants. All these procedures generate new $Q$-cohomologies.
Finally, multiplying these single-trace cohomologies generates multi-trace
cohomologies. At large $N$ where one can ignore the trace relations, these account
for all the supergravitons. At finite $N$, we shall still call them `graviton-like'
cohomologies as they do not give qualitatively new operators. At finite $N$,
their growths are slower than those of large $N$ gravitons due to the trace
relations between multi-trace operators.

Understanding where the non-graviton operators appear and how they look like
have been quite long standing problems. In particular, since the graviton-type
operators can be defined precisely at finite $N$, one can study this problem
from the simplest toy model of $SU(2)$ theory. Recently, \cite{Chang:2022mjp}
made a very promising discovery on this problem. They reported
the threshold level of the $SU(2)$ theory, from which such new operators exist. 
(They reported $7$ non-graviton states, but $6$ of them are descendants of 
the threshold state.)
The goal of this short note is to explicitly construct representatives of
such a cohomology and take a detailed look at their structures.\footnote{Although
not reported in \cite{Chang:2022mjp}, we believe that their raw data should also
contain all the information.} As we shall see below, one can find
a simple representative of  the cohomology.
It is our belief that the detailed shape of this operator
will inspire us towards better analytic studies,
e.g. ansatz for new cohomologies at higher $N$, etc.

\section{Results and discussions}

For convenience, we consider the $SU(2)$ super-Yang-Mills theory rather than
$U(2)$. The two theories are trivially related by multiplying
the free $U(1)$ theory. In the $SU(2)$ theory, \cite{Chang:2022mjp} reported
that the first non-graviton type cohomology appears at
$n\equiv 2\sum_{I=1}^3R_I+3\sum_{i=1}^2J_i=24$, at charges
$E=\frac{19}{2}$, $R_1=R_2=R_3=\frac{3}{2}$, $J_1=J_2=\frac{5}{2}$.
There is a unique operator in this sector, which is not of the graviton type. Furthermore, it was found that this operator is made of $7$ fields. 
Using these data, one can easily see from the charge table that 
only $\bar \phi^m, \psi_m, f$ can participate in the cohomology without 
any derivatives. Also, note from (\ref{Q-action}) that these three operators 
form a closed subsector under the $Q$-action. Even when checking 
the $Q$-exactness of an operator in this sector, $Q$ should act on 
another operator made only of $\bar\phi^m,\psi_m,f$ to be in this sector.
Thus, we
will not consider $\bar \lambda_{\dot \alpha}$ and $D_{+\dot\alpha}$ from now on.
By inspecting the charges, the operator may contain terms at
$f^0\psi^5\phi^2$, $f^1\psi^3\phi^3$ and $f^2\psi^1\phi^4$ orders.
One can further specify which $\psi$'s and $\bar\phi$'s can appear in which
combinations: this is helpful when listing all possible operators, but we skip
explaining this.

For the $SU(2)$ theory, the adjoint fields can be represented by three-dimensional vectors using the Pauli matrices $\left\{\frac{\sigma^{a=1,2,3}}{2}\right\}$ as the basis. Our vector notation of the BPS letters \eqref{BPS-letters} are given as follows:
\begin{equation}
g_{\textrm{YM}} \sqrt{2} \bar \phi^m \to \vec \phi_m\ ,  \;\; g_{\textrm{YM}} \psi_{m} \to \vec \psi_{m}\ , \;\;  -g_{\textrm{YM}} \sqrt{2}f \to \vec f\ .
\end{equation}
The $Q$-actions on the BPS letters \eqref{Q-action} in the vector notation are written as
\begin{equation}
Q \vec\phi_m = 0\ , \;\;
Q \vec\psi_{m} = \frac{1}{2} \epsilon_{mnp} \vec \phi_n  \times \vec \phi_p\ , \;\; Q \vec f = \vec\phi_{m} \times \vec \psi_m\ .
\end{equation}
For simplicity, we will drop $\;\vec{}\;$ symbol from all vectors from now on.
Also, we shall sometimes use the notation
$\phi_m=(\phi_1,\phi_2,\phi_3)\equiv (X,Y,Z)$.

\pagebreak
Writing down a cohomology is ambiguous by adding $Q$-exact terms, and
here we write down a simple representative. The unique cohomology in this sector may be represented as
\begin{eqnarray}\label{cohomology-rep1}
  O&\equiv&(\psi_1\cdot X-\psi_2\cdot Y)(\psi_3\cdot X)\psi_2\cdot(\psi_1\times \psi_1)
  +\textrm{cyclic}\\
  &\equiv&(\psi_1\cdot X-\psi_2\cdot Y)(\psi_3\cdot X)\psi_2\cdot(\psi_1\times \psi_1)
  +(\psi_2\cdot Y-\psi_3\cdot Z)(\psi_1\cdot Y)\psi_3\cdot(\psi_2\times \psi_2)
  \nonumber\\
  &&+(\psi_3\cdot Z-\psi_1\cdot X)(\psi_2\cdot Z)\psi_1\cdot(\psi_3\times \psi_3)\ ,
  \nonumber
\end{eqnarray}
where `cyclic' means adding two more terms obtained by making cyclic
permutations of $(X,\psi_1)$, $(Y,\psi_2)$ and $(Z,\psi_3)$.
The $Q$-closedness of this operator is easy to check. First note that
$\psi_3\cdot X$ and $\psi_1\cdot X-\psi_2\cdot Y$ are separately $Q$-closed.
In fact, they are graviton-type operators, for instance obtained by
\begin{equation}
  \psi_3\cdot X\sim [Q^2_+,X\cdot X]\sim [Q^1_+,X\cdot Y]\ ,\ \ 
  \psi_1\cdot X-\psi_2\cdot Y\sim [Q^3_+,X\cdot Y]  
\end{equation}
using the $SU(1,2|3)$
generators.
The $Q$-action on $\psi_2\cdot(\psi_1\times \psi_1)$ is given by
\begin{equation}\label{tr-rel}
  \{Q,\psi_2\cdot(\psi_1\times \psi_1)\}=2(Z\cdot\psi_1)(X\cdot \psi_1)
  +2(\psi_1\cdot Y)(\psi_2\cdot Z)-2(\psi_1\cdot Z)(\psi_2\cdot Y)\ .
\end{equation}
So the $Q$-action on the whole operator becomes
\begin{eqnarray}
  \{Q,O\}&=&-2(2Y)(3X)(1Z)(1X)
  +2(1X)(3X)(1Y)(2Z)-2(2Y)(3X)(1Y)(2Z)\nonumber\\
  &&-2(1X)(3X)(1Z)(2Z)+\textrm{cyclic}
  \nonumber\\
  &=&2(1X)(1Z)(2Y)(3X)+2(1X)(1Y)(2Z)(3X)-2(1X)(1Y)(2Z)(3X)\nonumber\\
  &&-2(1X)(1Z)(2Z)(3X)+\textrm{cyclic}=0\ ,
\end{eqnarray}
where $(1X)\equiv (\psi_1\cdot X)$, etc., and
the Fermi statistics forbids various terms like $(\psi_1\cdot X)^2$
or $(\psi_2\cdot Y)^2$. On the third line, we rearranged terms
using the cyclic permutation.

The fact that this is not $Q$-exact is harder to check. We comprehensively
studied all independent $Q$-exact operators in this sector, and have shown
that $O$ cannot be a linear combination of such operators.
To explain this, one should first write down all operators whose $Q$-actions
yield operators at the desired charges. Since we want $Q$-exact operators
made of $7$ letters, we should consider $Q$-action on
gauge-invariant operators made of $6$ fields among $f,\psi_m,\phi_m$. One can always
make $6$ vectors gauge-invariant by forming $3$ inner products. (For instance,
one may use an operator of the form $A\cdot (B\times C) \, D\cdot(E\times F)$, but
this can be rewritten as three inner products.) Since we
expect to find a unique cohomology \cite{Chang:2022mjp}, we can stay in
the sector invariant under the cyclic permutation symmetry, which is a symmetry 
of the $Q$-action. So
the $Q$-exact operators can also be restricted to cyclic invariant ones.
There exist 17 independent $Q$-exact and cyclic-invariant operators in this
sector.
They can be represented by the $Q$-action on the following 17 independent operators $O^\prime_{1,2,3, \cdots ,17}$:
\begin{align*}\label{17-exact}
    &O_1^\prime=( f \cdot  f) ( f \cdot  X) ( Y \cdot  Z)+\textrm{cyclic} \ , \\
    &O_2^\prime=( f \cdot  f) ( X \cdot  Y) ( \psi_1 \cdot  \psi_2)+\textrm{cyclic} \ , \\
    &O_3^\prime=( f \cdot  f) ( X \cdot  \psi_1) ( Y \cdot  \psi_2)+\textrm{cyclic} \ , \\
    &O_4^\prime=( f \cdot  X) ( f \cdot  Y) ( \psi_1 \cdot  \psi_2)+\textrm{cyclic} \ , \\
    &O_5^\prime=( f \cdot  \psi_1) ( f \cdot  \psi_2) ( X \cdot  Y)+\textrm{cyclic} \ , \\
    &O_6^\prime=( f \cdot  X) ( f \cdot  \psi_1) ( Y \cdot  \psi_2)+\textrm{cyclic} \ , \\
    &O_7^\prime=( f \cdot  Y) ( f \cdot  \psi_1) ( X \cdot  \psi_2)+\textrm{cyclic} \ , \\
    &O_8^\prime=( f \cdot  X) ( f \cdot \psi_2) ( Y \cdot  \psi_1)+\textrm{cyclic} \ , \\
    &O_{9}^\prime=( f \cdot  Y) ( f \cdot  \psi_2) ( X \cdot  \psi_1)+\textrm{cyclic} \ , \\
    &O_{10}^\prime=( f \cdot  f) ( X \cdot  \psi_2) ( Y \cdot  \psi_1)+\textrm{cyclic} \ , \\
    &O_{11}^\prime=( f \cdot  X) ( \psi_1 \cdot  \psi_2) ( \psi_1 \cdot  \psi_3)+\textrm{cyclic} \ , \\
    &O_{12}^\prime=( f \cdot  \psi_2) ( \psi_1 \cdot  X) ( \psi_1 \cdot  \psi_3)+\textrm{cyclic} \ , \\
    &O_{13}^\prime=( f \cdot  \psi_3) ( \psi_1 \cdot  \psi_2) ( \psi_1 \cdot  X)+\textrm{cyclic} \ , \\
    &O_{14}^\prime=( f \cdot  \psi_1) ( X \cdot  \psi_2) ( \psi_1 \cdot  \psi_3)+\textrm{cyclic} \ , \\
    &O_{15}^\prime=( f \cdot  X) ( f \cdot  \psi_1) ( X \cdot  \psi_1)+\textrm{cyclic} \ , \\
    &O_{16}^\prime=( f \cdot  \psi_1) ( \psi_2 \cdot  \psi_3) ( X \cdot  \psi_1)+\textrm{cyclic} \ , \\
    &O_{17}^\prime=( \psi_1 \cdot  \psi_2) ( \psi_2 \cdot  \psi_3) ( \psi_3 \cdot  \psi_1)\ .
    \stepcounter{equation}\tag{\theequation}
    \end{align*}
One can also consider cyclic-invariant $Q$-closed operators in this sector. Using
$7$ fields, one can always take gauge-invariant operators to take the form of
\begin{equation}\label{7-invariant}
  (A\cdot B)(C\cdot D)(E\cdot (F\times G))\ .
\end{equation}
There could be further relations between operators of this form, mostly due to the identity
\begin{equation}\label{vector-identity}
  A\otimes (B\times C)+B\otimes (C\times A)+C\otimes(A\times B)=
  [A\cdot (B\times C)]{\bf 1}_{3\times 3}\ ,
\end{equation}
which holds when $A,B,C$ are either: all bosonic, all fermionic, two bosonic and one fermionic. (Similar identity holds with signs changed when one field is bosonic and 
two are fermionic.) 
In any case, writing down all possible operators of the form (\ref{7-invariant}),
we checked on a computer all the relations between them and listed truly independent ones only.
Among them, we find $18$ independent $Q$-closed cyclic-invariant operators.
$17$ of them can be taken as the $Q$-actions of (\ref{17-exact}), and the remaining one
can be taken as (\ref{cohomology-rep1}). This proves that (\ref{cohomology-rep1}) can
be taken as the representative of the unique cohomology in this sector found in
\cite{Chang:2022mjp}.

Any linear combination between our cohomology operator (\ref{cohomology-rep1}) and the $Q$-exact operators given by the $Q$-actions on (\ref{17-exact}) will give an alternative representative of the cohomology operator. It would be nice if one can find other simple expressions.

We make several remarks and discussions on the future directions.

Beyond the threshold $n=24$, \cite{Chang:2022mjp} also found 
$6$ non-graviton cohomologies at $n=25$ order, with energies 
$E=10$, etc. These operators are essentially not new. They are 
the $\overline{Q}_{m\dot\alpha}$ superconformal descendants of 
the primary at $E=\frac{19}{2}$ that we have discussed so far.

We also note that our analysis was substantially simplified due to the 
fact that the $Q$-action closes within $\bar\phi^m,\psi_m,f$. This subsector cannot be reached just by specifying the charges 
$E,R_I,J_i$. For instance, the charges of $f$ are completely the same 
as those carried by two derivatives $D_{+\dot{+}}D_{+\dot{-}}$. 
What made us possible to restrict ourselves to this subsector is 
the `bonus symmetry' $Y$ \cite{Chang:2022mjp} of the cohomology problem, 
which is the number of fields in the operator. Restricting studies 
within this subsector, we think there is a good chance to make many 
parts of the calculations done in \cite{Chang:2022mjp} analytically.

We find that our representative (\ref{cohomology-rep1}) of the cohomology may show various interesting aspects. For instance, 
in (\ref{cohomology-rep1}), many terms after the $Q$-action vanish due to the Fermi statistics, such as 
$(\psi_1\cdot X)^2=0$. This aspect may be used to construct generalized 
ansatze for new cohomologies. It should be a subtle variant of the `Fermi surface' 
operators discussed in \cite{Berkooz:2006wc}. On the other hand, (\ref{cohomology-rep1}) might also  admit a `giant graviton' interpretation. That is to establish non-graviton cohomologies as determinant-like operators having open spin chains ending on them, 
generalizing the open giant magnons \cite{Hofman:2006xt}. Note that, inspired 
by the recent giant graviton rewriting of the index \cite{Imamura:2021ytr}, 
it was shown that the intersecting giant gravitons with open strings attached 
account for the black hole microstates \cite{Choi:2022ovw,Beccaria:2023hip,Kim:2024ucf}.

It would be also interesting to study the fate of our cohomology \eqref{cohomology-rep1} when $N>2$. For $N=2$, \eqref{cohomology-rep1} becomes $Q$-closed due to the trace relation \eqref{tr-rel} between $SU(2)$ matrices $(\psi_1 \cdot X),(\psi_1 \cdot Y),(\psi_1 \cdot Z),(\psi_2 \cdot Y),(\psi_2 \cdot Z)$, which does not hold for $SU(N)$ with $N>2$. This implies that \eqref{cohomology-rep1} will not be $Q$-closed for $N>2$. Explicitly, behavior of the 1-loop anomalous dimension of \eqref{cohomology-rep1} as varying $N$ was studied in \cite{Budzik:2023vtr}.

Recently, there has been various interesting progress on this subject, \textit{e.g.} \cite{Choi:2023znd,Chang:2023zqk,Choi:2023vdm,Chang:2024zqi}. In particular, \cite{Chang:2024zqi} classified the BPS operators into monotone and fortuitous ones based on their behaviors in the large $N$ limit. Our graviton operators correspond to the monotone operators, which form infinite sequences with increasing $N$, being $Q$-closed without using any trace relations of $SU(N)$. The non-graviton operators were called fortuitous operators, which become $Q$-closed by using the trace relations of $N \times N$ matrices valid only within the finite range of $N$.

\pagebreak

\hspace*{-0.8cm} {\bf\large Acknowledgements}
\vskip 0.2cm

\hspace*{-0.75cm} We thank Chi-Ming Chang for pointing out an error in the previous version.
This work is supported in part by a KIAS Individual Grant
PG081601, PG081602 at Korea Institute for Advanced Study (SC), the National Research
Foundation of Korea (NRF) Grants 2021R1A2C2012350 (SK, EL) and
the National Research
Foundation of Korea (NRF) Grants 2021R1A6A1A10042944, 2021R1A2C1012440 (JP).

\end{document}